\RequirePackage{lineno}
\documentclass[11pt,a4paper]{article}
\pdfoutput=1

\usepackage{jheppub}

\title{Limits on muon-neutrino to tau-neutrino oscillations induced by a sterile neutrino state obtained by OPERA at the CNGS beam}

\collaborationImg{\includegraphics[width=2.5cm]{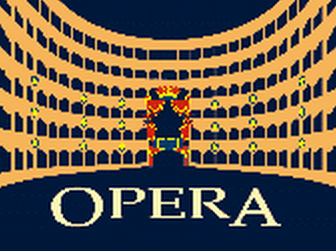}~~~The OPERA Collaboration}
\author[a]{N.~Agafonova,}
\author[b]{A.~Aleksandrov,}
\author[c]{A.~Anokhina,}
\author[d]{S.~Aoki,}
\author[e]{A.~Ariga,}
\author[e]{T.~Ariga,}
\author[f]{D.~Bender,}
\author[g]{A.~Bertolin,}
\author[h]{I.~Bodnarchuk,}
\author[i]{C.~Bozza,}
% ...

\affiliation[a]{INR - Institute for Nuclear Research of the Russian Academy of Sciences, RUS-117312 Moscow, Russia}
\affiliation[b]{INFN Sezione di Napoli, 80125 Napoli, Italy}
\affiliation[c]{ SINP MSU - Skobeltsyn Institute of Nuclear Physics, Lomonosov Moscow State University, RUS-119991 Moscow, Russia}
\affiliation[d]{ Kobe University, J-657-8501 Kobe, Japan}
\affiliation[e]{Albert Einstein Center for Fundamental Physics, Laboratory for High Energy Physics (LHEP), University of Bern, CH-3012 Bern, Switzerland }
\affiliation[f]{METU - Middle East Technical University, TR-06531 Ankara, Turkey}
\affiliation[g]{INFN Sezione di Padova, I-35131 Padova, Italy}
\affiliation[h]{JINR - Joint Institute for Nuclear Research, RUS-141980 Dubna, Russia}
\affiliation[i]{Dipartimento di Fisica dell'Universit\`a di Salerno and ``Gruppo Collegato'' INFN, I-84084 Fisciano (Salerno), Italy}
% ...

\emailAdd{alessandro.paoloni@lnf.infn.it}
\emailAdd{alessandra.pastore@ba.infn.it}
\emailAdd{antonia.dicrescenzo@na.infn.it}

\abstract{The OPERA experiment, exposed to the CERN to Gran Sasso $\nu_\mu$
beam, collected data from 2008 to 2012. Four oscillated $\nu_\tau$
Charged Current interaction candidates have been detected in
appearance mode, which are consistent with $\nu_\mu \to \nu_\tau$
oscillations at the atmospheric $\Delta m^2$ within the
``standard'' three-neutrino framework. In this paper, the OPERA $\nu_\tau$ 
appearance results are used to derive limits on the mixing parameters of a 
massive sterile neutrino.}


\arxivnumber{1503.01876}

\begin{document} 
\maketitle
\flushbottom

\section{Introduction}

The OPERA experiment \cite{opera} operated in the CERN Neutrinos
to Gran Sasso (CNGS) beam produced at CERN and directed towards
the Gran Sasso Underground Laboratory of INFN (LNGS), 730 km away,
where the detector is located. The experiment is unique in its
capability to observe $\nu_\tau$ appearance on an event-by-event basis. 
Nuclear emulsion films instrumenting the target allow the detection of the 
short-lived $\tau$ lepton decay, and hence the identification of
$\nu_\tau$ Charged Current (CC) interactions. The standard three-neutrino 
oscillation framework predicts $\nu_\mu \to \nu_\tau$
oscillations with close-to-maximal mixing at the so-called
atmospheric scale, $\Delta m^2_{32} \sim 2.4 \times 10^{-3}$ $\mathrm {eV^2}$
\cite{pdg}, i.e. in the oscillation parameters region discovered
by detecting atmospheric neutrinos \cite{atm}. OPERA has observed
four $\nu_\tau$ CC interaction candidate events \cite{tau1}
\cite{tau2} \cite{tau3} \cite{tau4}, consistent with the
expectation of the standard oscillation framework at this scale. This result represents the first direct
evidence of $\nu_\mu \to \nu_\tau$ oscillation in appearance mode.

In the present paper, limits are derived on the existence of a
massive sterile neutrino. The excess of $\nu_e$ ($\overline{\nu}_e$)
observed by the LSND \cite{lsnd} and MiniBooNE
\cite{miniboone} collaborations and the so-called
reactor \cite{reactor} and Gallium \cite{gallium} neutrino
anomalies are also interpreted as due to the existence of a fourth 
sterile neutrino with mass at the eV scale.  
In relation to this issue, it is worth mentioning that the effective number of neutrino-like species decoupled from the primeval plasma measured by the Planck collaboration is 3.15 $\pm$ 0.23 at 95\% Confidence Level (CL) \cite{bplanck}.

Neutrino oscillations at large $\Delta m^2$ have been
searched for by several short baseline experiments. The most stringent
limits on $\nu_\mu \to \nu_\tau$ oscillations were set by the NOMAD
\cite{nomad} and CHORUS \cite{chorus} experiments, with high
sensitivity for  $\Delta m^2$ values larger than 10 $\mathrm {eV^2}$.

In the following, a short description of the OPERA experimental
setup and of the procedure used to detect $\nu_\tau$ interactions
is given, the data analysis is described and exclusion regions in the parameter space are
derived.

\section{Detector, beam, and data sample }

In OPERA, CNGS neutrinos interacted in a massive target made of
lead plates interspaced with nuclear emulsion films acting as high
accuracy tracking devices \cite{Detector2}. This kind of detector
is historically called \emph{Emulsion Cloud Chamber} (ECC). The
OPERA detector is made of two identical Super Modules, each
consisting of a target section and a muon magnetic spectrometer.
Each target section has a mass of about 625 tons and is made of
$\sim$75000 units called \emph{bricks}. A target brick consists of 56
lead plates 1mm-thick interleaved with 57 nuclear emulsion films
300 $\mu$m-thick for a mass of 8.3 kg. Its total thickness along the
beam direction corresponds to about 10 radiation lengths. The
bricks are assembled in vertical walls instrumented with
scintillator strips (\emph{Target Tracker} detectors, TT) to trigger the
read-out and locate neutrino interactions within the target.
\begin{figure}[h]
\centering
\begin{center}
\includegraphics[width=4.0in]{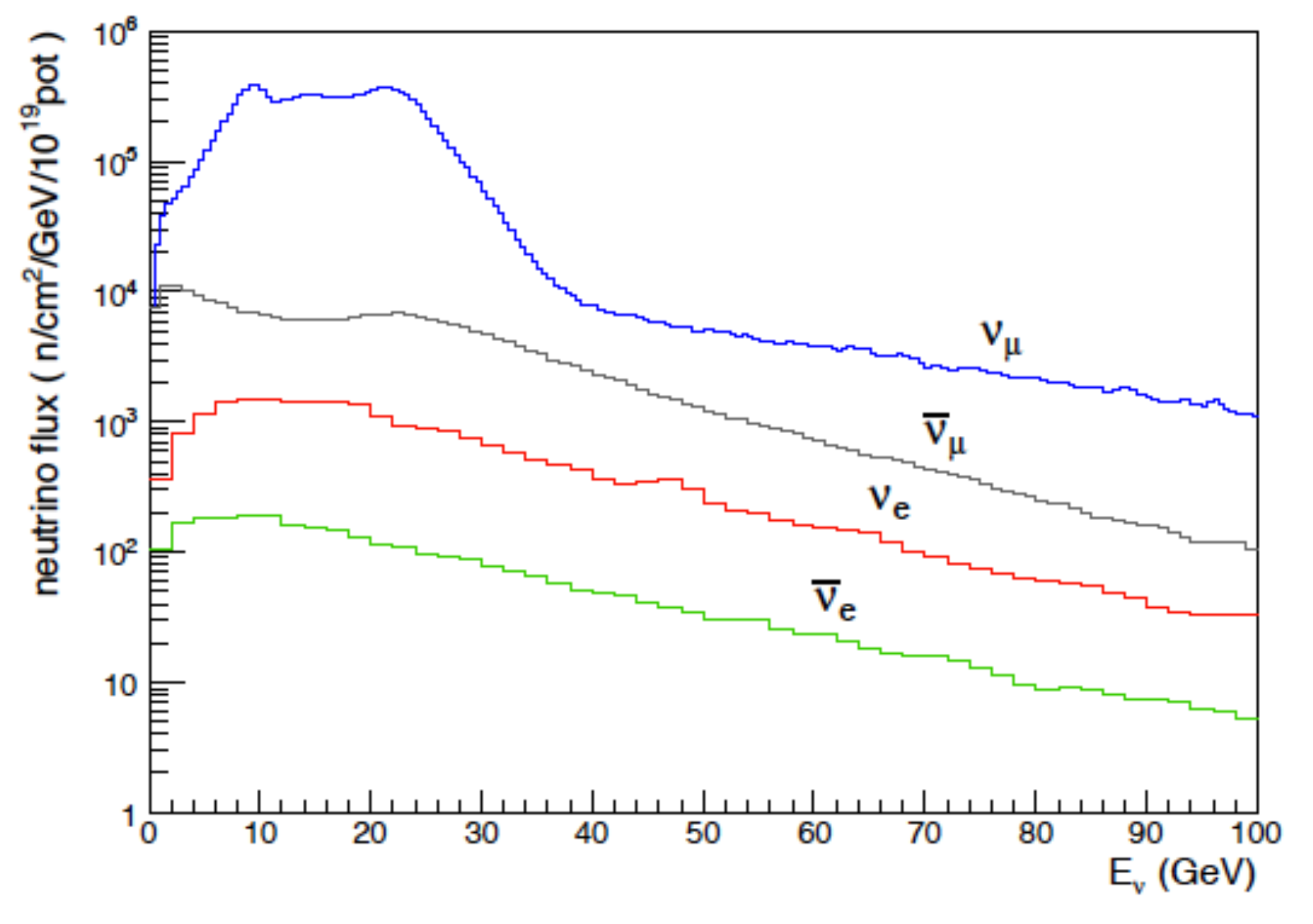}
\end{center}
\caption{\label{fig:flux} Fluxes of the different CNGS beam neutrino components
at LNGS.}
\end{figure}

OPERA was exposed to the  CNGS $\nu_\mu$ beam generated by protons from the SPS
accelerator at CERN \cite{beam}.  The contaminations of
$\overline{\nu}_\mu$, $\nu_e$ and $\overline{\nu}_e$ CC
interactions at LNGS, relative to $\nu_\mu$ CC interactions, are 2.1\%, 0.9\% and less than 0.1\%, respectively. The contamination
of prompt $\nu_\tau$ is negligible. The average $\nu_\mu$ energy
is 17 GeV. The energy spectra of the four beam components at the
detector site are shown in figure \ref{fig:flux} \cite{beamflux}.

The data taking, based on a minimum bias interaction trigger from
the TT scintillators, started in 2008 and ended in
December 2012. OPERA collected data corresponding to $17.97 \times
10^{19}$ protons on target (pot) with 19505 recorded events. The
data sample used in this analysis is defined following the selection
criteria described in \cite{tau2} and corresponds to about 75\% of
the total statistics.

 \section{Search for $\nu_\tau$ interactions}

Bricks selected as candidates to contain CNGS neutrino
interactions are analysed following the procedure described in
detail in \cite{tau2}. Here we just recall the main steps of the
analysis.

The brick where a neutrino interaction occurred is predicted 
by the electronic detectors and extracted from the target
by an automatic brick manipulator system. Two extra low background
emulsion films (\emph{Changeable Sheets}, CS)  \cite{CS} located
downstream of the brick act as an interface between the brick
and the electronic detectors. If the measurement of the CS yields
tracks related to the neutrino interaction, the emulsion films of
the brick are developed. Their analysis provides the three
dimensional reconstruction of the neutrino interaction and of
the possible decay vertices of short-lived particles
\cite{decaysearch} with micrometric accuracy.

This procedure has led to the detection of four $\nu_\tau$ CC
interaction candidates. The total expected background in the
analysed sample amounts to $0.23 \pm 0.05$ events. The absence of
a $\nu_\mu \to \nu_\tau$ oscillation signal, i.e. the hypothesis of the four events being
background, is excluded with a significance of 4.2 $\sigma$
\cite{tau4}.

\section{Sterile neutrino search via $\nu_\mu \to \nu_\tau$ oscillations}

In \cite{tau4} the detection of four $\nu_\tau$ CC events is compared to the 
expectation for $\nu_\mu \to \nu_\tau$ oscillations in the atmospheric sector, 
computed within a simplified two-flavour scheme assuming full mixing and 
$|\Delta m^2_{32}| = 2.32 \times 10^{-3}$ $\mathrm {eV^2}$ \cite{pdgold}.
The expected number of events is $2.30 \pm 0.46$ ($2.21 \pm 0.44$) assuming 
normal (inverted) hierarchy of neutrino masses; the number is obtained by 
rescaling the value given in \cite{tau4} for 
$|\Delta m^2_{32}| \approx |\Delta m^2_{31}| \approx |\Delta m^2| = 2.43 
\times 10^{-3}$ $\mathrm {eV^2}$
($|\Delta m^2| = 2.38 \times 10^{-3}$ $\mathrm {eV^2}$) \cite{pdg},
where $\Delta m^2$ is defined as $m_3^2 - \frac{(m_1^2+m_2^2)}{2}$. 
By including the background, $2.53 \pm 0.46$ ($2.44 \pm 0.44$) events are 
expected in total. The error, which is dominated by the uncertainty on the $\tau$ detection 
efficiency and on the $\nu_{\tau}$ interaction cross section, also takes into
account the experimental precision on the atmospheric oscillation
parameters.
The observation of four events is compatible with these expectations. 
Despite the limited statistics, an excess or a deficit of $\nu_\tau$ interactions due to $\nu_\mu \to \nu_\tau$ oscillations induced by the mixing with a sterile neutrino can be  evaluated.

In presence of a fourth sterile neutrino with mass $m_4$, the
oscillation probability is a function of the 4 $\times$ 4 mixing
matrix \textit{U} and of the three squared mass differences.
Defining $C=2|U_{\mu 3}||U_{\tau 3}|$, $\Delta_{ij}=1.27~\Delta m^2_{ij}~L/E$ 
($i$,$j$ = 1,2,3,4), $\phi_{\mu \tau}=~Arg(U_{\mu 3}U^*_{\tau 3}U^*_{\mu 4}U_{\tau 4})$ 
and $\sin 2\theta_{\mu \tau}=2 |U_{\mu 4}| |U_{\tau 4}|$, 
the $\nu_{\mu} \rightarrow \nu_{\tau}$ oscillation probability \textit{P(E)} 
can be parametrised as:

\begin{equation}
\begin{split}
P(E)~&=~C^2~\sin^2 \Delta_{31} ~+~
\sin^2 2\theta_{\mu \tau} ~\sin^2 \Delta_{41} \\
&+~ \frac{1}{2}~C~\sin2\theta_{\mu \tau}~\cos\phi_{\mu \tau}~
\sin 2 \Delta_{31} ~\sin 2 \Delta_{41}\\
&-~ C~\sin 2\theta_{\mu \tau}~\sin \phi_{\mu \tau}~
\sin^2 \Delta_{31} ~\sin 2 \Delta_{41} \\
&+~ 2~C~\sin 2\theta_{\mu \tau} ~\cos \phi_{\mu \tau} ~
\sin^2 \Delta_{31} ~\sin^2 \Delta_{41}\\
&+~ C~\sin 2\theta_{\mu \tau} ~ \sin \phi_{\mu \tau} ~
\sin 2 \Delta_{31}  ~\sin^2 \Delta_{41} 
\end{split}
\label{eq1}
\end{equation}
where $\Delta m^2_{31}$ and $\Delta m^2_{41}$ are expressed in
$\mathrm {eV^2}$, $L$ in km and $E$ in GeV. Given the long baseline and
the average CNGS neutrino energy, \textit{P(E)} is independent of
$\Delta m^2_{21}$, since $\Delta_{21} \approx 4 \times 10^{-3}$.
The
terms proportional to $\sin \phi_{\mu \tau}$ are CP-violating, while those
proportional to $\sin 2 \Delta_{31}$ are sensitive to the
mass hierarchy of the three standard neutrinos, normal ($\Delta
m^2_{31}>0$) or inverted ($\Delta m^2_{31}<0$). 
Matter effects have been checked to be negligible for 
$\Delta m^2_{41}>$ 1 $\mathrm {eV^2}$.

Observed neutrino oscillation anomalies \cite{bform}, if 
interpreted in terms of one additional sterile neutrino, suggest
$|\Delta m^2_{41}|$ values at the $\mathrm {eV^2}$ scale (the so-called 
3+1 model).
In the following, unless stated otherwise, the analysis will be restricted
only to positive $\Delta m^2_{41}$ values, since negative values are disfavoured
by results on the sum of neutrino masses from cosmological surveys 
\cite{bplanck}.
For $\Delta m^2_{41}>$ 1 $\mathrm {eV^2}$, at the concerned domain of $L/E$
 and taking into account the finite energy resolution,
$\sin 2~\Delta_{41}$ and $\sin^2 \Delta_{41}$ average to 0 and $\frac{1}{2}$, 
respectively. 
The oscillation probability $P(E)$ can thus be approximated to \cite{bform}:

\begin{equation}
\label{eappr1}
\begin{split}
P(E)~&=~C^2~\sin^2 \Delta_{31}~+~ \frac{1}{2} ~\sin^2 2\theta_{\mu \tau}\\
&+~C ~\sin 2\theta_{\mu \tau} ~\cos \phi_{\mu \tau}~\sin^2 \Delta_{31}\\
&+~ \frac{1}{2} ~C ~\sin 2\theta_{\mu \tau} ~\sin \phi_{\mu \tau} ~\sin 2 \Delta_{31}.
\end{split}
\end{equation}

In order to obtain an upper limit on $\sin^2 2\theta_{\mu \tau}$
at high values of $\Delta m^2_{41}$, the likelihood is defined as
$\it{L}$($\phi_{\mu \tau}$, $ \sin^2 2\theta_{\mu \tau}$, $ C^2$)$ = e^{-\mu}~\mu
^n/n!$, where $n=4$ is the number of $\nu_{\tau}$ candidate events
and $\mu$ is the expected number of events, $\mu~=~n_b~+~A~\int 
\phi(E) P(E) \sigma(E) \epsilon(E) ~ dE$. $n_b=0.23$ is the
expected number of background events \cite{tau4}, $\phi(E)$ is 
the $\nu_\mu$ flux shown in figure 1, $P(E)$ is the oscillation probability 
given by equations \eqref{eq1} or \eqref{eappr1}, $\sigma (E)$ is the  
$\nu_{\tau}$ CC interaction cross section, $\epsilon (E)$ is the $\tau$
detection efficiency and $A$ is a 
normalisation factor proportional to the fraction of the analysed sample 
and to the target mass.

The analysis presented here is based on the asymptotic $\chi^2$ 
distribution of the log likelihood ratio test statistics:
$ q~=-2~\ln(\widetilde{\it{L}}(\phi_{\mu \tau} ,\sin^2 2\theta_{\mu \tau})/\it{L_0})$,
where $\it{L_0}=e^{-n}~n ^n/n!$ and
$\widetilde{\it{L}}(\phi_{\mu \tau} ,\sin^2 2\theta_{\mu \tau})$ is the profile
likelihood obtained by maximising $\it{L}$($\phi_{\mu \tau}$ ,$ \sin^2 2\theta_{\mu \tau}$, $ C^2$) over $C^2$.
By definition, $C^2$  ranges between 0 and 1, but for any pair of values of $\sin^2 2\theta_{\mu \tau}$ and 
$\phi_{\mu \tau}$, it is limited by the unitarity of the 
mixing matrix; the likelihood is maximised accordingly.
The value of $|\Delta m^2_{31}|$ has been fixed to 
$2.43 \times 10^{-3}$ $\mathrm {eV^2}$ for the normal hierarchy and to
$2.38 \times 10^{-3}$ $\mathrm {eV^2}$ for the inverted hierarchy of the
three standard neutrinos \cite{pdg}.

\begin{figure}[h!]
\centering
\begin{center}
\includegraphics[width=6.0in]{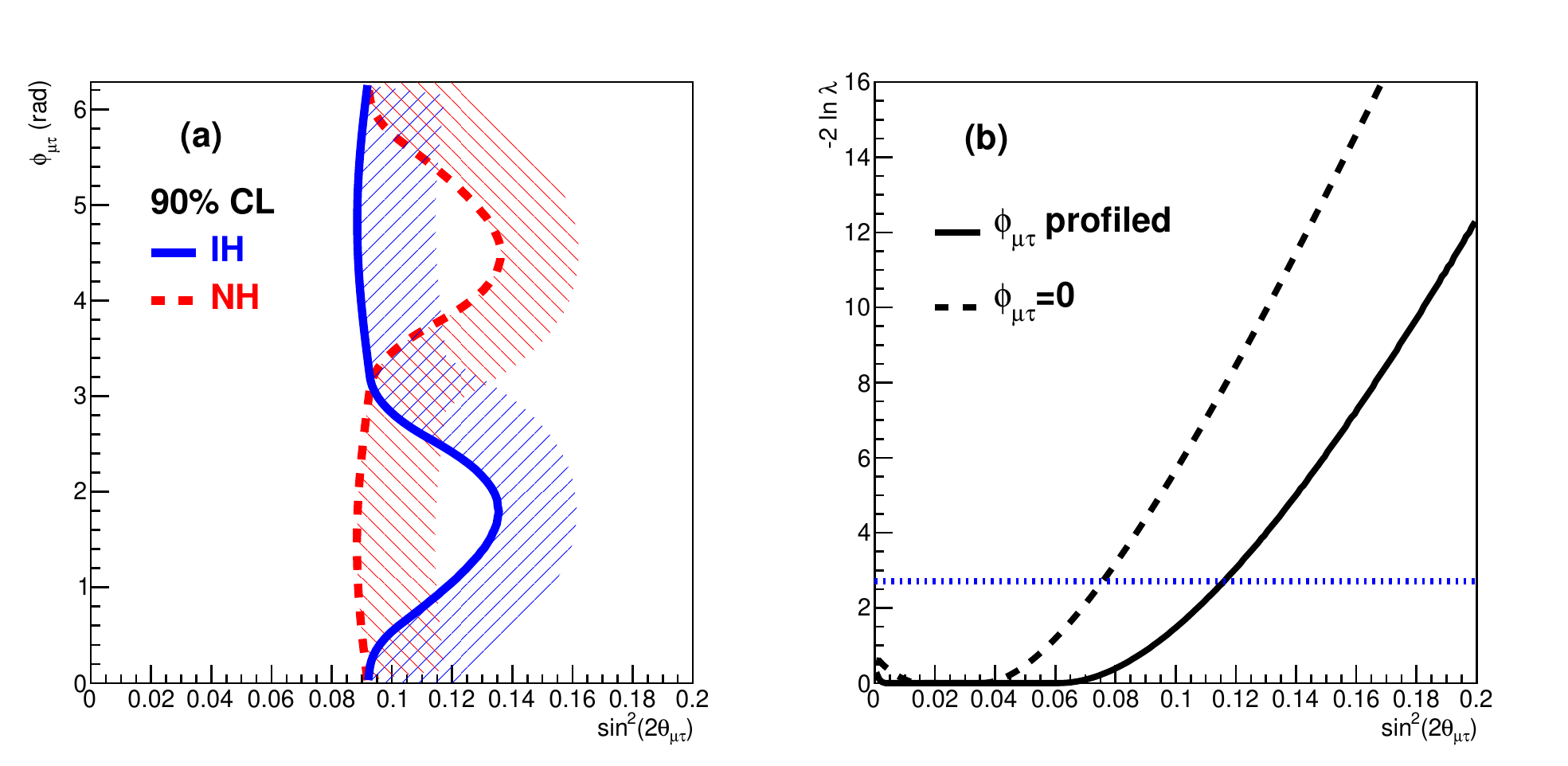}
\end{center}
\caption{\label{fig:plot1} (a) $90\%$ CL exclusion limits in the
$\phi_{\mu \tau}$ vs $\sin^2 2\theta_{\mu \tau}$ parameter space 
for normal (NH, dashed red) and inverted (IH, solid blue) hierarchies
assuming $\Delta m^2_{41}>$ 1 $\mathrm{eV^2}$. 
Bands are drawn to indicate the excluded regions.
(b) Log likelihood ratio as a function of $\sin^2 2\theta_{\mu \tau}$
for $\phi_{\mu \tau}=0$ (dashed line) and for the profile likelihood (continuous
line).
}
\end{figure}
\begin{figure}[h!]
\centering
\begin{center}
\includegraphics[width=4.0in]{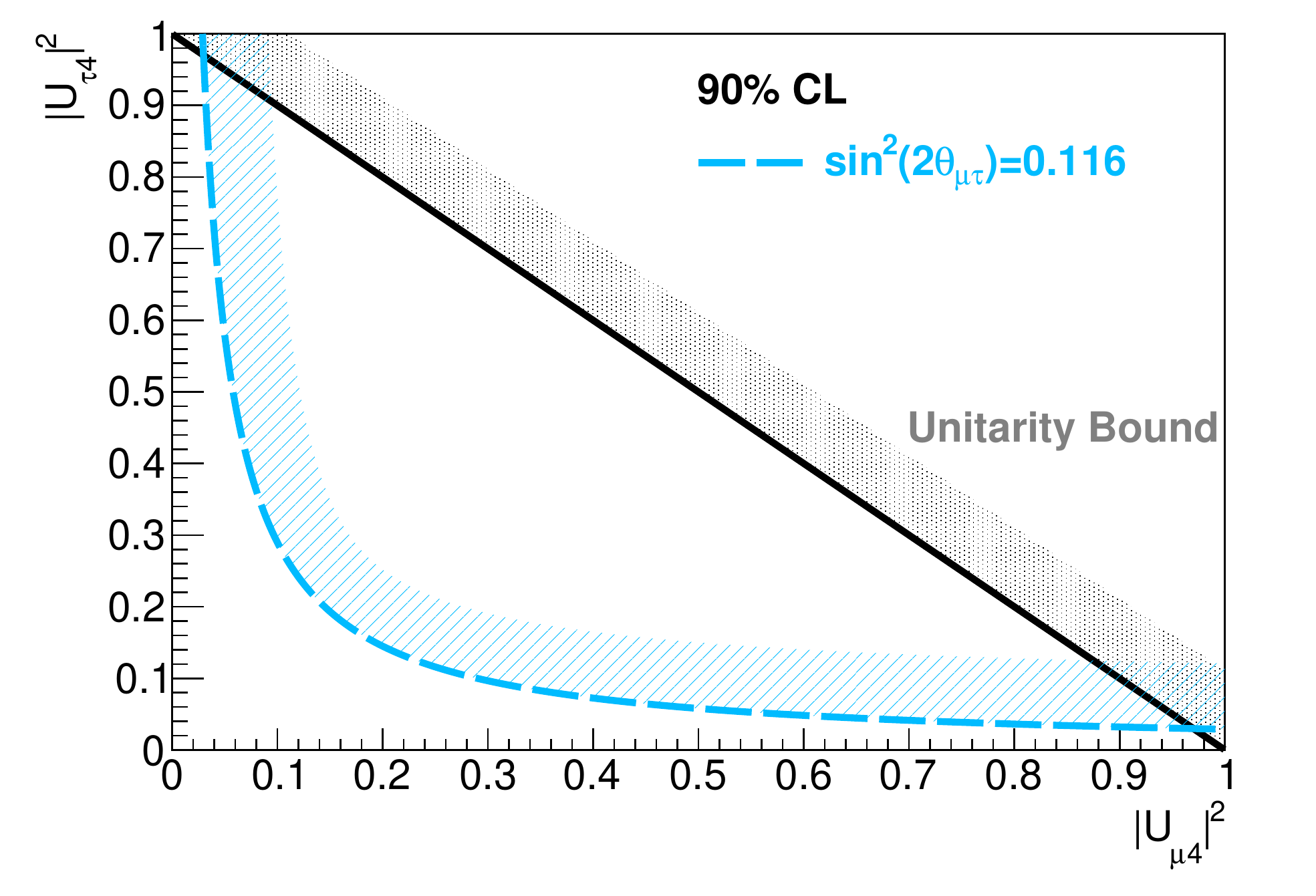}
\end{center}
\caption{\label{fig:plot2} $90\%$ CL exclusion limits (blue line)
in the $|U_{\tau 4}|^2$ vs $|U_{\mu 4}|^2$ plane 
assuming $\Delta m^2_{41}>$ 1 $\mathrm {eV^2}$.
The unitarity bound (grey line) is also shown. 
Bands are drawn to indicate the excluded regions. 
}
\end{figure}

In figure \ref{fig:plot1}(a) the $90\%$ CL exclusion
limits are presented for both normal and inverted mass hierarchies in
the parameter space of $\phi_{\mu \tau}$ vs $\sin^2 2\theta_{\mu \tau}$. 
The edge of the excluded region ranges from 0.088 to 0.136 in
$\sin^2 2\theta_{\mu \tau}$ for both mass hierarchies of the three standard
neutrinos. 
For any fixed value of $\phi_{\mu \tau}$, $q$ is distributed according
to a $\chi^2$ statistics with one degree of freedom.
Profiling the likelihood also over $\phi_{\mu \tau}$, as shown in figure
\ref{fig:plot1}(b), an upper limit of 0.116 is obtained at $90\%$ CL on 
$\sin^2 2\theta_{\mu \tau}$, almost independently of the hierarchy of the 
three standard neutrino masses.
A negligible difference arises from the different $|\Delta m^2_{31}|$ value
used in the analysis.
The $\sin^2 2\theta_{\mu \tau}$ upper limit is affected by a $20\%$ systematic 
error from the uncertainties on the $\tau$ detection efficiency and
$\nu_{\tau}$ interaction cross section.

Given the definition of $\sin^2 2\theta_{\mu \tau}$ in terms of
$U_{\mu 4}$ and $U_{\tau 4}$, it is possible to translate the
upper limit on  $\sin^2 2\theta_{\mu \tau}$ into an exclusion
curve in the $|U_{\mu4}|^2$ vs $|U_{\tau4}|^2$ plane, as shown in figure 
\ref{fig:plot2} 
together with the unitarity bound
($|U_{\mu4}|^2+|U_{\tau4}|^2\le1$).

\begin{figure}[h]
\centering
\begin{center}
\includegraphics[width=3.5in]{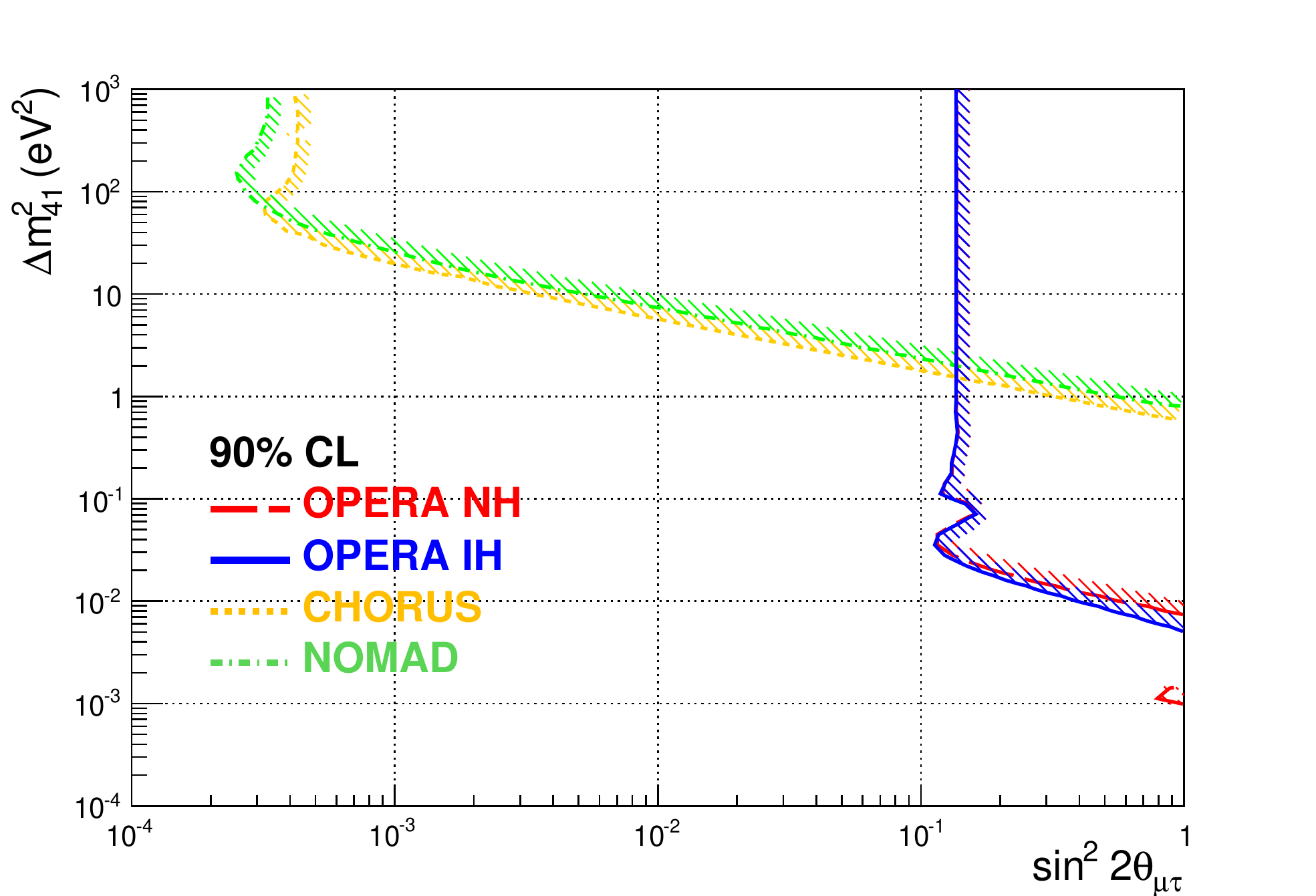}
\end{center}
\caption{\label{fig:exclusion} OPERA 90\% CL exclusion limits
in the $\Delta m^2_{41}$ vs $\sin^2 2\theta_{\mu \tau}$ parameter space 
for the normal (NH, dashed red) and inverted (IH, solid blue) hierarchy of
the three standard neutrino masses. The exclusion plots by NOMAD \cite{nomad} 
and CHORUS \cite{chorus} are also shown. Bands are drawn to indicate the 
excluded regions.}
\end{figure}

To extend the search for a possible fourth sterile neutrino down to
small $\Delta m^2_{41}$ values, the likelihood has been computed using
the GLoBES software \cite{bglobes}, in order to take into account
also matter effects and the non-zero value of $\Delta m^2_{21}$.
The likelihood has been profiled also on the $\Delta m^2_{31}$ value.
More details on the analysis are available in \cite{noteglobes}. 
In figure \ref{fig:exclusion} the 90\% CL exclusion plot is reported in 
the $\Delta m^2_{41}$ vs $\sin^2 2\theta_{\mu \tau}$ parameter space.
The most stringent limits of direct searches for 
$\nu_\mu \to \nu_\tau$ oscillations at short-baselines obtained
by the NOMAD \cite{nomad} and CHORUS \cite{chorus} experiments are also
shown.
Our analysis stretches the limits on $\Delta m^2_{41}$ down to 10$^{-2}~\mathrm {eV^2}$, extending the values explored with the $\tau$ appearance searches by about two orders of magnitude at large mixing, for $\sin^2 2\theta_{\mu\tau} \gtrsim ~0.5$.
For maximal mixing, the $90\%$ CL excluded region extends down to 
$\Delta m^2_{41}$ = 7.4 (5.2) $\times 10^{-3}~\mathrm {eV^2}$ for normal
(inverted) hierarchy of the three standard neutrino masses, with a $10\%$
systematic error deriving from the uncertainties on the $\tau$ detection 
efficiency and $\nu_{\tau}$ interaction cross section.

A narrow region is excluded at $90\%$ CL at 
$\Delta m^2_{41} \approx 10^{-3}~\mathrm {eV^2}$ for the normal hierarchy
of the three standard neutrinos. It arises from a suppression
of the $\nu_\mu \to \nu_\tau$ oscillation probability due to the presence of 
the sterile neutrino. 
Instead, the oscillation probability is enhanced at full mixing and 
high $\Delta m^2_{41}$ values.
For a number of $\tau$ neutrino candidates equal to the expectation in the three neutrino framework, the excluded region at $\Delta m^2_{41} \approx 10^{-3}~\mathrm {eV^2}$ would disappear.

The analysis was performed assuming $\Delta m^2_{41}>0$.
Since present limits on the sum of neutrino masses from cosmological surveys
do not exclude small negative values for $\Delta m^2_{41}$, the analysis was repeated following this assumption.
The exclusion plots obtained in this way are similar to
those of figure \ref{fig:exclusion}, but with hierarchies exchanged.
It is worth underlining that the results obtained in
the 3+1 model, shown in figures \ref{fig:plot1} and \ref{fig:plot2}, are 
independent of the sign of $\Delta m^2_{41}$, as is the probability in 
equation \ref{eappr1}.

Assuming $CP$ conservation, that implies $\sin \phi_{\mu \tau} = 0$, and neglecting terms in $\sin^2 \Delta_{31}$ in equation \eqref{eq1}, of the order of $10^{-2}$ at CNGS energies $E_{CNGS}$, the oscillation probability at high values of $\Delta m^2_{41}$ approximates to that of a two-flavour model parametrised in terms of two effective mixing parameters, $\theta_{eff}$ and $\Delta m^2_{eff}$:
\begin{equation*}
\lim_{E \rightarrow E_{CNGS}} P(E|CP~\mathrm {conservation}) \approx \sin^2 2\theta_{eff} \cdot sin^2 (1.27 ~\Delta m^2_{eff}~ L/E) 
\label{eq2nu}
\end{equation*}
In this framework, with 4 observed events and 2.53 $\pm$ 0.46 events expected from the normal hierarchy of standard oscillations, including background, the upper limit on the number of additional $\nu_\tau$ events, evaluated in the Feldman-Cousins approach \cite{feld}, is 6.4 at 90\% CL. The upper limit on $\sin^2 2\theta_{eff}$  is 0.069, to be compared with 0.076 for $\sin^2 2\theta_{\mu \tau}$ at $\phi_{\mu \tau}=0$ (see figure \ref{fig:plot1}(b)).

\section{Conclusions}

The OPERA experiment was designed to observe $\nu_\mu \to
\nu_\tau$ oscillations through $\nu_\tau$ appearance at a
baseline of 730 km in the CNGS beam. Exploiting its unique
capability to identify $\tau$ neutrino interactions, OPERA has
observed four $\nu_\tau$ CC candidate interactions, consistent with the expected number of oscillation events in the standard three-neutrino framework,  
2.30 $\pm$ 0.46 (2.21 $\pm$ 0.44), for the normal (inverted) mass hierarchy and 0.23 $\pm$ 0.05 background events.

In this paper we present limits on the existence of a sterile neutrino in 
the 3+1 neutrino model. 
At high values of $\Delta m^2_{41}$, 
the measured $90\%$ CL upper limit on the mixing term 
$\sin^2 2\theta_{\mu \tau}=4 |U_{\mu 4}|^2 |U_{\tau 4}|^2$ is 0.116,
independently of the mass hierarchy of the three standard neutrinos.
The OPERA experiment extends the exclusion limits on $\Delta m^2_{41}$
in the $\nu_\mu \to \nu_\tau$ appearance channel down to values
of $10^{-2}~\mathrm {eV^2}$ at large mixing for $\sin^2 2\theta_{\mu \tau} \gtrsim ~0.5$.

\acknowledgments
We thank CERN for the successful operation of the CNGS facility and INFN for the continuous support given to the experiment through its LNGS laboratory. We acknowledge funding from our national agencies: Fonds de la Recherche Scientique-FNRS and Institut InterUniversitaire des Sciences Nucleaires
for Belgium, MoSES for Croatia, CNRS and IN2P3 for France, BMBF for Germany, INFN for Italy, JSPS
(Japan Society for the Promotion of Science), MEXT
(Ministry of Education, Culture, Sports, Science and
Technology), QFPU (Global COE programme of Nagoya University, Quest for Fundamental Principles in the Universe supported by JSPS and MEXT) and Promotion and Mutual Aid Corporation for Private Schools of Japan for Japan, SNF, the University of Bern for Switzerland, the Russian Foundation for Basic Research (grant no. 19-02-00213-a, 12-02-12142 ofim), the Programs of the Presidium of the Russian Academy
of Sciences Neutrino physics and Experimental and theoretical researches of fundamental interactions connected
with work on the accelerator of CERN, the Programs
of Support of Leading Schools (grant no. 3110.2014.2),
and the Ministry of Education and Science of the Russian
Federation for Russia and the National Research Foundation
of Korea Grant (NRF-2013R1A1A2061654) for Korea.
We are also indebted to INFN for providing fellowships and grants to non-Italian researchers. We thank the IN2P3 Computing Centre (CC-IN2P3) for providing computing resources for the analysis and hosting the central database for the OPERA experiment.
We are indebted to our technical collaborators for the excellent quality of their work over many years of design, prototyping, construction and running of the detector and of its facilities.
We also want to thank Carlo Giunti, Francesco Vissani and Enrico Nardi for fruitful discussions.

\end{document}